\documentclass[Letter, longauth, utf8]{aa}  

\usepackage{graphicx}
\usepackage{txfonts}
\usepackage[colorlinks=true,citecolor=blue, urlcolor=blue, linkcolor=blue]{hyperref}
\newcommand{\change}[1]{\textcolor{black}{#1}}
\newcommand{\changetwo}[1]{\textcolor{black}{#1}}

\begin{document} 

   \title{Using VLTI/GRAVITY+ to determine the identity of a third planet candidate in the PDS 70 system}

   \author{D.~Trevascus\inst{\ref{MPIA}, \ref{Heidelberg}},
          W.~Brandner\inst{\ref{MPIA}}, 
          O.~Balsalobre-Ruza\inst{\ref{CAB}, \ref{UCM}},
          S.~Lacour\inst{\ref{LIRA}},
          K.~Abd El Dayem\inst{\ref{LIRA}},
          N.~Aimar\inst{\ref{FEUP},\ref{CENTRA}},
          A.~Berdeu\inst{\ref{LIRA},\ref{ESOGarching}},
          J.-P.~Berger\inst{\ref{IPAG}},        
          G.~Bourdarot\inst{\ref{MPE}},
          V.~Christiaens\inst{\ref{CEASaclay}, \ref{STARLiege}, \ref{KUL}},
          C.~Correia\inst{\ref{FEUP},\ref{CENTRA}},      
          R.~Davies\inst{\ref{MPE}},
          D.~Defr{\`e}re\inst{\ref{KUL}},
          A.~Drescher\inst{\ref{MPE}},                  
          A.~Eckart\inst{\ref{UCologne},\ref{MPIfRA}},        
          F.~Eisenhauer\inst{\ref{MPE},\ref{TUM}},
          M.~Fabricius\inst{\ref{MPE}},
          H.~Feuchtgruber\inst{\ref{MPE}},       
          S.~Flesch\inst{\ref{MPE}},
          N.M.~F{\"o}rster~Schreiber\inst{\ref{MPE}},
          A.~Foschi\inst{\ref{LIRA}},        
          Q.~Fournier\inst{\ref{MPE}},
          P.~Garcia\inst{\ref{FEUP},\ref{CENTRA}},        
          R.~Garcia~Lopez\inst{\ref{UCD}},
          R.~Genzel\inst{\ref{MPE},\ref{UBerkley}},        
          S.~Gillessen\inst{\ref{MPE}},       
          I.~Hammond\inst{\ref{MPIA}},
          S.F.~H{\"o}nig\inst{\ref{USouthampton}},      
          M.~Houll{\'e}\inst{\ref{IPAG}},
          S.~Joharle\inst{\ref{MPE}},
          P.~Kervella\inst{\ref{LIRA}},                          
          L.~Kreidberg\inst{\ref{MPIA}},
          L. Labadie\inst{\ref{UCologne}},
          O.~Lai\inst{\ref{CotedAzur}},
          R.~Laugier\inst{\ref{KUL}},
          J.-B.~Le~Bouquin\inst{\ref{IPAG}},
          J.~Leftley\inst{\ref{USouthampton}},
          R.~Li\inst{\ref{MPE}},
          B.~Lopez\inst{\ref{CotedAzur}},
          D.~Lutz\inst{\ref{MPE}},     
          G.-D.~Marleau\inst{\ref{UDE},\ref{MPIA},\ref{Bern}},
          F.~Mang\inst{\ref{MPE}},
          A.~M{\'e}rand\inst{\ref{ESOGarching}},
          F.~Millour\inst{\ref{CotedAzur}},
          M.~Montarg{\`e}s\inst{\ref{LIRA}},           
          N.~Moruj{\~a}o\inst{\ref{FEUP},\ref{CENTRA}},
          H.~Nowacki\inst{\ref{CotedAzur}},
          M.~Nowak\inst{\ref{LIRA}},
          J.~Osorno\inst{\ref{LIRA}},
          T.~Ott\inst{\ref{MPE}},           
          S.~Pappert\inst{\ref{MPE}},
          T.~Paumard\inst{\ref{LIRA}},
          K.~Perraut\inst{\ref{IPAG}}, 
          G.~Perrin\inst{\ref{LIRA}},
          R.~Petrov\inst{\ref{CotedAzur}},
          P.O.~Petrucci\inst{\ref{IPAG}},        
          N.~Pourr{\'e}\inst{\ref{IPAG}},           
          S.~Rabien\inst{\ref{MPE}},
          D.C.~Ribeiro\inst{\ref{MPE}},           
          S.~Robbe-Dubois\inst{\ref{CotedAzur}},
          M.~Sadun~Bordoni\inst{\ref{MPE}},
          J.~Sanchez~Bermudez\inst{\ref{UNAM}},
          D.~Santos\inst{\ref{MPE}}, 
          J.~Sauter\inst{\ref{MPIA}, \ref{Heidelberg}},
          J.~Scigliuto\inst{\ref{CotedAzur}},  
          J.~Shangguan\inst{\ref{Kavli}},   
          T.T.~Shimizu\inst{\ref{MPE}},   
          F.~Soulez\inst{\ref{CRAL}},
          C.~Straubmeier\inst{\ref{UCologne}},
          E.~Sturm\inst{\ref{MPE}},   
          M.~Subroweit\inst{\ref{UCologne}},   
          C.~Sykes\inst{\ref{USouthampton}},
          L.J.~Tacconi\inst{\ref{MPE}},   
          P.~Th{\'e}venet\inst{\ref{LIRA}},
          I.~Urso\inst{\ref{LIRA}},
          F.~Vincent\inst{\ref{LIRA}},   
          J.~Woillez\inst{\ref{ESOGarching}},
          and the GRAVITY+ Collaboration
          }

        \institute{Max Planck Institute for Astronomy, K{\"o}nigstuhl 17, 69117 Heidelberg, Germany
        \label{MPIA}
        \and
        Universität Heidelberg, Grabengasse 1, 69117 Heidelberg, Germany
        \label{Heidelberg}
        \and
        Centro de Astrobiología (CAB), CSIC-INTA, ESAC campus, Camino Bajo del Castillo s/n, 28692, Villanueva de la Cañada (Madrid), Spain
        \label{CAB}
        \and
        Departamento de Física de la Tierra y Astrofísica, Facultad de Ciencias Físicas, Universidad Complutense de Madrid, 28040 Madrid, Spain
        \label{UCM}
        \and
        LIRA, Observatoire de Paris, Universit{\'e} PSL, CNRS, Sorbonne Universit{\' e}, Universit{\' e} de Paris, 5 place Jules Janssen, 92195 Meudon, France
        \label{LIRA}
        \and
        Faculdade de Engenharia, Universidade do Porto, rua Dr. Roberto Frias, 4200-465 Porto, Portugal
        \label{FEUP}
        \and
        CENTRA - Centro de Astrof{\' i}sica e Gravita\c{c}{\~a}o, IST, Universidade de Lisboa, 1049-001 Lisboa, Portugal
        \label{CENTRA}
        \and
        European Southern Observatory, Karl-Schwarzschild-Stra{\ss}e 2, 85748 Garching, Germany
        \label{ESOGarching}
        \and
        Univ. Grenoble Alpes, CNRS, IPAG, 38000 Grenoble, France
        \label{IPAG}
        \and
        Max Planck Institute for Extraterrestrial Physics, Giessenbachstra{\ss}e 1, 85748 Garching, Germany
        \label{MPE}
        \and
        Université Paris-Saclay, Université Paris Cité, CEA, CNRS, AIM, F-91191 Gif-sur-Yvette, France
        \label{CEASaclay}
        \and
        STAR Institute, Universit\'e de Li\`ege, All\'ee du Six Ao\^ut 19c, 4000 Li\`ege, Belgium
        \label{STARLiege}
        \and
        Institute of Astronomy, KU Leuven, Celestijnenlaan 200D, 3001, Leuven, Belgium
        \label{KUL}
        \and
        1st Institute of Physics, University of Cologne, Z{\"u}lpicher Straße 77, 50937 Cologne, Germany
        \label{UCologne}
        \and
        Max Planck Institute for Radio Astronomy, auf dem H{\"u}gel 69, 53121 Bonn, Germany
        \label{MPIfRA}
        \and
        Department of Physics, Technical University of Munich, 85748 Garching, Germany
        \label{TUM}
        \and
        School of Physics, University College Dublin, Belfield, Dublin 4, Ireland
        \label{UCD}
        \and
        Departments of Physics \& Astronomy, Le Conte Hall, University of California, Berkeley, CA 94720, USA
        \label{UBerkley}
        \and
        School of Physics \& Astronomy, University of Southampton, Southampton, SO17 1BJ, United Kingdom
        \label{USouthampton}
        \and
        Universit{\'e} C{\^o}te d{'}Azur, Observatoire de la C{\^o}te  d{'}Azur, CNRS, Laboratoire Lagrange, France
        \label{CotedAzur}
        \and
        Fakult\"at für Physik, Universit\"at Duisburg--Essen, Lotharstra\ss{}e 1, 47057 Duisburg, Germany
        \label{UDE}
        \and
        Division of Space Research \& Planetary Sciences, Physics Institute, University of Bern, Gesellschaftsstr.~6, 3012 Bern, Switzerland%
        \label{Bern}
        \and
        Universidad Nacional Aut{\'o}noma de M{\'e}xico, Instituto de Astronom{\'i}a, A. P. 702-64, 04510, Ciudad de M{\'e}xico, M{\'e}xico
        \label{UNAM}
        \and
        Univ. Lyon, Univ. Lyon 1, ENS de Lyon, CNRS, Centre de Recherche Astrophysique de Lyon UMR5574, F-69230, Saint Genis-Laval, France
        \label{CRAL}
        \and
        The Kavli Institute for Astronomy and Astrophysics, Peking University, Beijing 100871, China
        \label{Kavli}
        }

   \date{Received May 1, 2026; accepted June 13, 2026}

 
  \abstract
   {Detections of protoplanets are rare and protoplanetary disk features mischaracterized as planets are common. PDS 70 is \changetwo{one of only two stars} known to host multiple confirmed protoplanets, PDS 70 b and c, and repeat detections of a third point-like source in the system suggest the presence of third inner planet. However, previous observations of this third source are insufficient to distinguish whether it is a planet or a concentrated dust clump in Keplerian motion. Our observations with VLTI/GRAVITY+ did not re-detect this point-like source, suggesting that it is, in fact, a dust clump and not a planet. These observations demonstrate how the angular resolving power of VLTI/GRAVITY+ can be used to distinguish between protoplanets and protoplanetary disk features.}

    \authorrunning{Trevascus et al.}
    \maketitle
    
\section{Introduction}
Protoplanets are a common explanation for the ubiquity of substructure in protoplanetary disks \citep[e.g.,][]{Dong_2015, Zhang_2018}. Despite several large surveys \citep{Currie_2019, Xie_2020, Zurlo_2020, Follette_2023}, protoplanets have only been successfully detected in a handful of systems, including PDS 70 \citep{Keppler_2018, Haffert_2019}, AB Aurigae \citep{Currie_2022}, HD 169142 \citep{Hammond_2023}, WISPIT 2 \changetwo{\citep{Close_2025, vanCapelleveen_2025, Lawlor_2026}}, and 2MASS J16120668-3010270 \citep{Li_2025}.

PDS 70 is host to two known protoplanets, PDS 70 b and c, and it is currently \changetwo{one of only two stars} around which multiple protoplanets have been confirmed. These two planets were detected in H$\alpha$ \citep{Haffert_2019}, indicating ongoing accretion, and their near-infrared photometry is consistent with thermal emission \citep{Keppler_2018, Mesa_2019, Wang_2021, Blakely_2025}.

Multiple papers have reported the detection of a third potential planet in the PDS 70 system. \cite{Mesa_2019} observed a point-like source interior to PDS 70 b and c to the northwest of the star with VLT/SPHERE. \cite{Christiaens_2024} redetected this source with JWST/NIRCam. This source also appeared in re-reductions of archival VLT/SPHERE and VLT/SINFONI data spanning 10 years by \cite{Hammond_2025}. However, the current observations of this source are insufficient to differentiate if it is a true protoplanet or only a concentrated dust clump. The source has not been detected in H$\alpha$, and its infrared photometry more closely matches starlight reflected off dust than thermal emission from a planet \citep{Hammond_2025}.

For most current telescopes, observation of the third planet candidate is challenging since its separation from its host star is small ($\sim 100$ mas). The GRAVITY instrument on the Very Large Telescope Interferometer (VLTI) is therefore well placed to search for the inner planets given its ability to produce sub-milliarcsecond angular resolution observations in the $K$ band \citep[e.g.,][]{Nowak_2020, Wang_2021}. The recent GRAVITY+ upgrade to this instrument, consisting of an upgrade to the adaptive optics system, is designed to improve these qualities even further, pushing the instrument to higher contrasts at short separations \citep{GRAVITY+_2025}.

In this letter we describe our attempt to observe this third planet candidate with VLTI/GRAVITY+ and the subsequent non-detection. In Section \ref{sec:obs} we describe our observations and how we reduced the observational data. In Section \ref{sec:discussion} we discuss the implications of our non-detection and state our conclusions in Section \ref{sec:conclusions}. 

\section{Observations and data reduction}
\label{sec:obs}

\subsection{GRAVITY+ observations}
\label{sec:obs_gravity}

We observed PDS 70 on 16 February 2025 with the GRAVITY+ instrument using the four 8m Unit Telescopes (UTs) of ESO’s VLTI. The star PDS 70 A itself served as the visual natural guide star for the adaptive optics and as the $K$-band fringe tracking reference source. We collected data using the $K$-band medium resolution mode ($R \approx 500$). The observing setup we used followed the strategy outlined in \cite{Nowak_2020} and \cite{Wang_2021}, where integrations with the science fiber set on the central star are placed between planet integrations. Table~\ref{tab:obs_log} gives an overview of the integration times, atmospheric conditions, and coherence time ($\tau_0$).

\begin{table*}
\caption{Observing log of the new GRAVITY+ astrometric measurements reported in this paper.}
\begin{centering}
    \label{tab:obs_log}
    \begin{tabular}{lcccccr}
         \hline \hline
Date & MJD & DIT/NDIT/NEXP & Airmass & Seeing (arcsec) & $\tau_0$ (ms) & ESO Program \\
\hline
2025-02-16 & 60722 & 100s/4/4	& 1.045--1.072 & 0.37--1.42 & 9.0--12.1 & 0114.C-0147 \\
\hline
\end{tabular}
\end{centering}
\end{table*}

\begin{figure*}
    \centering
    \includegraphics[width=0.75\linewidth]{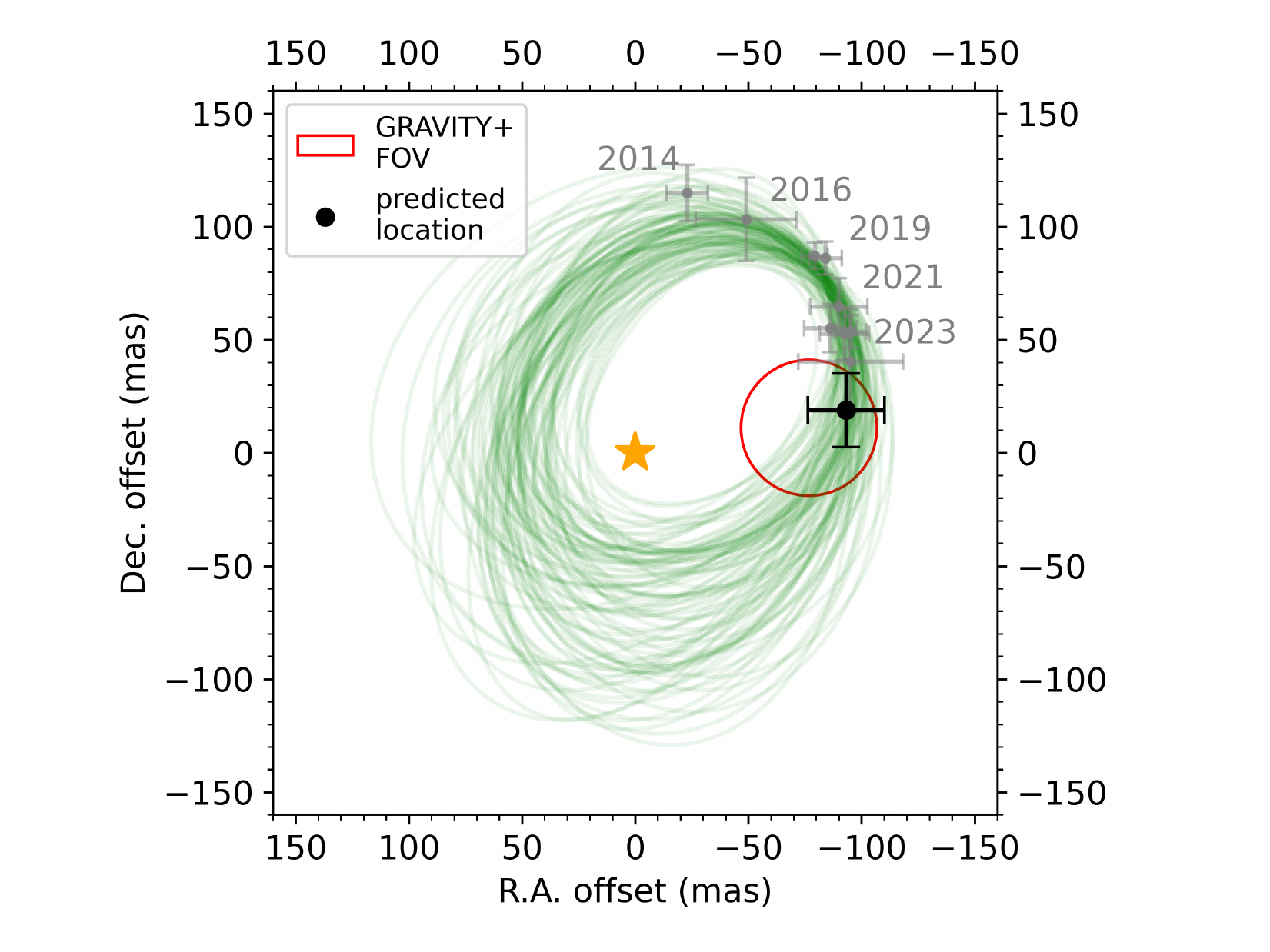}
    \caption{VLTI/GRAVITY+ FOV (red circle) for our attempted observation of the inner planet candidate in the PDS 70 system plotted alongside the potential orbits of the candidate (in green). The gray points indicate the literature astrometry for the candidate, \changetwo{while the black dot shows the predicted position of the candidate on the night of observation, with error bars showing the 2$\sigma$ error on this prediction}.
    }
    \label{fig:PDS70d_obs}
\end{figure*}

\changetwo{
Figure~\ref{fig:PDS70d_obs} shows the predicted location of the candidate on the night of our observation relative to the GRAVITY field of view (FOV) and alongside some of the potential orbits for the candidate. The predicted position of the candidate is $(\Delta{\rm RA},~\Delta{\rm{Dec}})~=~(-93\pm17,~19\pm16)$, with the errors given as the $2\sigma$ uncertainty.
}

\subsection{Data reduction}
\label{sec:data_reduction}

The data were reduced using Public Release 1.7.0 of the ESO GRAVITY pipeline \citep{GRAVITY_DRS2014}. They were reduced up until the ''astroreduced'' intermediate data products, in which individual integrations are not averaged.

Subsequent analysis used the \verb<exoGRAVITY< pipeline, following a similar procedure to the one described in Appendix B of \cite{Nowak_2020}. Equation~B.4 of \cite{Nowak_2020} describes the on-planet visibility ($V_{\rm model}$) for a planet observed by GRAVITY, which depends on the relative RA ($\Delta\alpha$) and Dec. ($\Delta\delta)$ from the star. For any given set of GRAVITY data with visibilities $V_{\rm data}$, a $\chi^2$ sum can be calculated to find the likelihood that the data corresponds to planet emission:
\begin{equation}
    \chi^2_{\rm planet}(\Delta\alpha,\Delta\delta) = [V_{\rm data} - V_{\rm model}]^T W^{-1} [V_{\rm data} - V_{\rm model}],
\end{equation}
where $W$ is the corresponding covariance matrix. A $\chi^2$ map can be generated by calculating $\chi^2$ for a set of $(\Delta\alpha,\Delta\delta)$ points in the GRAVITY FOV.

If we take $(\Delta\alpha, \Delta\delta)=(0,0)$, this corresponds to a model with no planet, which we can use as a reference. By comparing the planet and no planet models, we can define
\begin{equation}
    z(\Delta\alpha, \Delta\delta) = \chi^2_{\rm no\,planet} - \chi^2_{\rm planet}(\Delta\alpha, \Delta\delta).
    \label{eq:z}
\end{equation}
We note that $z(\Delta\alpha, \Delta\delta)$ compares the likelihood of models with and without a planet at the location $(\Delta\alpha, \Delta\delta)$. 

\begin{figure*}
    \centering
    \includegraphics[width=0.75\linewidth]{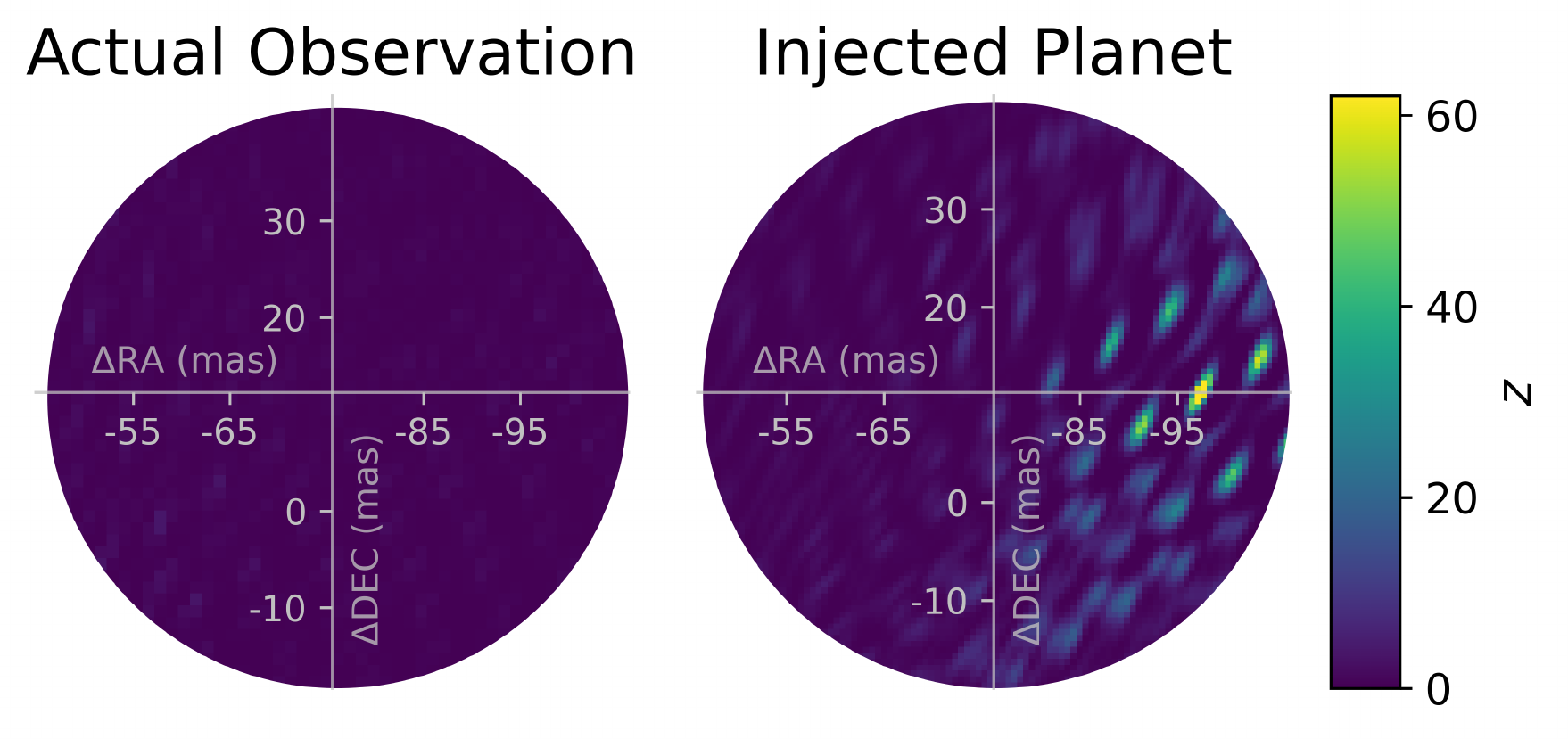}
    \caption{Maps of $z$ for our observation with VLTI/GRAVITY+. The left panels show our observation alone. The right panels show our observation with a fake planet signal injected at the predicted position of the third planet candidate. The contrast of the injected signal matches the K-band contrast of the candidate given by \cite{Hammond_2025}.}
    \label{fig:z_maps}
\end{figure*}

Figure~\ref{fig:z_maps} shows the map of $z$ for the observation described in this letter. Both panels show the results from our observation, but for the right panel, a fake planet signal was injected into the visibility data at the expected position for the planet candidate. The injected signal takes the form of Equation~B.4 from \cite{Nowak_2020} with a contrast of $8.4$ magnitudes, matching the K-band contrast for the planet candidate given by \cite{Hammond_2025}. From these $z$ maps we determined that there is no planet signal in our attempted observation of the third planet candidate.

\section{Discussion}
\label{sec:discussion}

\change{
Our non-detection of the candidate suggests that it is not a planet and is instead a concentrated dust clump. The existing $K$-band photometry of the candidate \citep{Hammond_2025} indicates that it has a contrast similar to PDS 70 c. However, PDS 70 c appears as a point source after the data reduction process described in Section~\ref{sec:data_reduction} \citep{Wang_2021, Trevascus_2025}, while this candidate does not. This is also consistent with the spectrum of the candidate, which resembles reflected starlight more than thermal emission \citep{Hammond_2025}.
}

\change{
The key difference between our observations and previous detections of this source \citep[e.g., with VLT/SPHERE][]{Mesa_2019, Hammond_2025} is the increased resolving power granted via the long baseline of GRAVITY ($\sim 100$~m). GRAVITY has a nominal angular resolution of $2$ mas, corresponding to $\sim 0.2$~au at the distance of PDS 70. Any significantly smaller object, such as a Jupiter-sized planet ($2R_{\rm Jup}=0.001$~au), should therefore have high visibility and appear as a point-like source. A source of size $\sim 0.2$~au or larger, such as a dust clump, may appear point-like for other instruments but not for GRAVITY.
}

\change{
The visibility model described in Section~\ref{sec:data_reduction} corresponds specifically to a point-like source and therefore is not suitable for the detection of extended sources, such as dust clumps. From our analysis alone we are only able to put a rough lower limit on the extent of the source, and we leave further visibility modeling to future work.
}

\change{It is still possible that the GRAVITY science fiber was not well positioned to image the third planet candidate, especially given that a pointing error of $30$ mas or more can result in a significant loss of flux \citep{Perrin_2019}}. Orbit modeling from \cite{Trevascus_2025} was used to predict the position of the planet candidate on the night of observation. However, the observation was performed before the orbit modeling was finalized and therefore the FOV does not cover the outermost orbits as shown in Figure~\ref{fig:PDS70d_obs}. \change{The predicted location of the planet based on its median stable orbit is still situated within the FOV, although it is separated from the center of the FOV by 22.5~mas. We also used the range of stable orbit parameters for the planet candidate from \cite{Trevascus_2025} to determine the probability that the candidate was outside the FOV on the night of the observation. We found that 12.6 percent of the possible positions lay outside the FOV.}

\subsection{Trojan dust}

\cite{Blakely_2025} also did not detect a third planet candidate to the northwest of PDS 70 when observing it with the Aperture Masking Interferometric mode of JWST/NIRISS. They did, however, detect excess emission to the south of the star that was not co-located with PDS 70 b or c.

Recent multiwavelength VLT/SPHERE observations by \cite{Balsalobre-Ruza_2026} identified additional emission with a dust-like spectrum at a similar separation from the star as the point source detected by \cite{Mesa_2019}, \cite{Christiaens_2024}, and \cite{Hammond_2025} but located to the southeast of the star. Their follow-up VLTI/GRAVITY+ observations resulted in the tentative detection of a point source with a position angle compatible with the JWST/NIRISS excess detected by \cite{Blakely_2025}.

\cite{Balsalobre-Ruza_2026} proposed that both the southeast and northwest emissions are Trojan dust collected within the L4 and L5 Lagrangian points of a new candidate inner planet, matching theoretical simulations \citep{Montesinos_2020}. All the new detected sources lie along the orbit predicted by \cite{Hammond_2025} for a potential third inner planet corresponding to the northwest emission.

It is possible that the proximity of this new planet candidate to the inner disk is what causes the Trojan dust in its orbit to be an especially bright source. The inner disk is a plentiful source of small dust grains that may be easily captured at the L4 and L5 Lagrange points \citep{Jang_2024, Pinilla_2024}. Closer to the star, individual dust grains are also able to scatter more stellar flux. In contrast, Trojan dust in the orbit of PDS 70 b has been tentatively detected with ALMA \citep{Balsalobre-Ruza_2023} but is yet to be identified in the near-infrared.

\subsection{Differentiating between planets and dust substructure}

PDS 70 is not the first directly imaged system in which emission sources have been misidentified as planet candidates. \cite{Kalas_2008} observed an optical point source orbiting the star Formalhaut that they characterized as a planet candidate. However the candidate was not detected in the near-infrared \citep{Marengo_2009}, and follow-up observations by \cite{Gaspar_2020} observed that the source had extended in size and faded in brightness. This source is now characterized as a dust cloud created after the collision of two planetesimals due to the detection of a similar event in the same system \citep{Kalas_2026}.

A similar case study exists for the young ($\sim 2$ Myr) sun-like star LkCa 15. \cite{Kraus_2012} and \cite{Sallum_2015} claimed the detection of multiple protoplanets orbiting this star from observations in the near-infrared. However follow-up observations by \cite{Currie_2019} and \cite{Sallum_2023} identified these signals as parts of the inner dust disk.

As evidenced by the observations described in this letter, interferometry can be a useful tool to differentiate between true planets and dust disk substructure. Therefore instruments such as VLTI/GRAVITY+ are particularly powerful for identifying planets at small separations.

\section{Conclusions}
\label{sec:conclusions}

We attempted to observe the candidate third planet in the PDS 70 system with VLTI/GRAVITY+. Our observations did not detect point source emission at the expected location of this candidate. Therefore we conclude that the previously detected emission sources correspond to a dust clump instead of a planet.

\begin{acknowledgements}
We thank Cade B{\"u}rgy for useful discussions. 

GDM acknowledges the support of the Deutsche Forschungsgemeinschaft (DFG) through grant MA~9185/2-1.

SFH acknowledges support through UK Research and Innovation (UKRI) under the UK government’s Horizon Europe Funding Guarantee (EP/Z533920/1, selected in the 2023 ERC Advanced Grant round) and an STFC Small Award (ST/Y001656/1).

J.S-B. acknowledges the support received by the UNAM DGAPA-PAPIIT project AG-101025 and from the SECIHTI Ciencia de Frontera project CBF-2025-I-3033.

We are very grateful to our funding agencies (MPG, ERC, CNRS [PNCG, PNGRAM], DFG, BMBF/BMFTR, Paris Observatory [CS, PhyFOG], Observatoire des Sciences de l'Univers de Grenoble, and the Funda\c c\~ao para a Ci\^encia e a Tecnologia), to ESO and the Paranal staff, and to the many scientific and technical staff members in our institutions, who helped to make GRAVITY/GRAVITY+ a reality.
\end{acknowledgements}

\bibliographystyle{yahapj}
\bibliography{ref}

\begin{appendix}
\end{appendix}

\end{document}